\documentclass[aps,prc,twocolumn,superscriptaddress,nofootinbib]{revtex4-1}

\usepackage{float}
\usepackage{graphicx}
\usepackage[T1]{fontenc}

\begin{document}

\title{Shell evolution approaching the $N=20$ island of inversion: Structure of $^{29}$Mg}


\author{A. Matta}
\email[Corresponding author:]{matta@lpccaen.in2p3.fr}
\affiliation{Department of Physics, University of Surrey, Guildford GU2 5XH, UK}
\affiliation{Laboratoire de Physique Corpusculaire, 14000  Caen, France}
\author{W.N. Catford}
\affiliation{Department of Physics, University of Surrey, Guildford GU2 5XH, UK}
\author{N.A. Orr}
\affiliation{Laboratoire de Physique Corpusculaire, 14000  Caen, France}
\author{J. Henderson}
\affiliation{TRIUMF, 4004 Wesbrook Mall, Vancouver, British Columbia, V6T 2A3, Canada}
\author{P. Ruotsalainen}
\affiliation{TRIUMF, 4004 Wesbrook Mall, Vancouver, British Columbia, V6T 2A3, Canada}
\author{G. Hackman}
\affiliation{TRIUMF, 4004 Wesbrook Mall, Vancouver, British Columbia, V6T 2A3, Canada}
\author{A. B. Garnsworthy}
\affiliation{TRIUMF, 4004 Wesbrook Mall, Vancouver, British Columbia, V6T 2A3, Canada}
\author{F. Delaunay}
\affiliation{Laboratoire de Physique Corpusculaire, 14000  Caen, France}
\author{R. Wilkinson}
\affiliation{Department of Physics, University of Surrey, Guildford GU2 5XH, UK}
\author{G. Lotay}
\affiliation{Department of Physics, University of Surrey, Guildford GU2 5XH, UK}
\author{Naofumi Tsunoda}
\affiliation{Center for Nuclear Study, the University of Tokyo, 7-3-1 Hongo, Bunkyo-ku, Tokyo, Japan}
\author{Takaharu Otsuka}
\affiliation{Center for Nuclear Study, the University of Tokyo, 7-3-1 Hongo, Bunkyo-ku, Tokyo, Japan}
\affiliation{Department of Physics and Center for Nuclear Study,The University of Tokyo, 7-3-1 Hongo, Bunkyo-ku, Tokyo, Japan}
\author{A.J. Knapton}
\affiliation{Department of Physics, University of Surrey, Guildford GU2 5XH, UK}
\author{G. C. Ball}
\affiliation{TRIUMF, 4004 Wesbrook Mall, Vancouver, British Columbia, V6T 2A3, Canada}
\author{N. Bernier}
\affiliation{Department of Physics and Astronomy, University of British Columbia, Vancouver, British Columbia V6T 1Z1, Canada }
\affiliation{TRIUMF, 4004 Wesbrook Mall, Vancouver, British Columbia, V6T 2A3, Canada}
\author{C. Burbadge}
\affiliation{Department of Physics, University of Guelph, 50 Stone Road East, Guelph, Ontario  N1G 2W1, Canada}
\author{A. Chester}
\affiliation{Department of Chemistry, Simon Fraser University, 8888 University Drive, Burnaby, British Columbia V5A 1S6, Canada}
\author{D. S. Cross}
\affiliation{Department of Chemistry, Simon Fraser University, 8888 University Drive, Burnaby, British Columbia V5A 1S6, Canada}
\author{S. Cruz}
\affiliation{Department of Physics and Astronomy, University of British Columbia, Vancouver, British Columbia V6T 1Z1, Canada }
\affiliation{TRIUMF, 4004 Wesbrook Mall, Vancouver, British Columbia, V6T 2A3, Canada}
\author{C. Aa. Diget}
\affiliation{Department of Physics, University of York, York, YO10 5DD, UK}
\author{T. Domingo}
\affiliation{Department of Chemistry, Simon Fraser University, 8888 University Drive, Burnaby, British Columbia V5A 1S6, Canada}
\author{T.E. Drake}
\affiliation{Department of Physics, University of Toronto, Toronto, Ontario M5S 1A7, Canada}
\author{L.J. Evitts}
\affiliation{TRIUMF, 4004 Wesbrook Mall, Vancouver, British Columbia, V6T 2A3, Canada}
\affiliation{Department of Physics, University of Surrey, Guildford GU2 5XH, UK}
\author{F.H. Garcia}
\affiliation{Department of Chemistry, Simon Fraser University, 8888 University Drive, Burnaby, British Columbia V5A 1S6, Canada}
\author{S. Hallam}
\affiliation{Department of Physics, University of Surrey, Guildford GU2 5XH, UK}
\affiliation{TRIUMF, 4004 Wesbrook Mall, Vancouver, British Columbia, V6T 2A3, Canada}
\author{E. MacConnachie}
\affiliation{TRIUMF, 4004 Wesbrook Mall, Vancouver, British Columbia, V6T 2A3, Canada}
\author{M. Moukaddam}
\affiliation{Department of Physics, University of Surrey, Guildford GU2 5XH, UK}
\affiliation{TRIUMF, 4004 Wesbrook Mall, Vancouver, British Columbia, V6T 2A3, Canada}
\author{D. Muecher}
\affiliation{Department of Physics, University of Guelph, 50 Stone Road East, Guelph, Ontario  N1G 2W1, Canada}
\author{E. Padilla-Rodal}
\affiliation{Instituto de Ciencias Nucleares, Universidad Nacional Aut\'onoma de M\'exico, AP 70-543, Mexico City 04510, DF, Mexico}
\author{O. Paetkau}
\affiliation{TRIUMF, 4004 Wesbrook Mall, Vancouver, British Columbia, V6T 2A3, Canada}
\author{J. Park}
\affiliation{Department of Physics and Astronomy, University of British Columbia, Vancouver, British Columbia V6T 1Z1, Canada }
\affiliation{TRIUMF, 4004 Wesbrook Mall, Vancouver, British Columbia, V6T 2A3, Canada}
\author{J.L. Pore}
\affiliation{Department of Chemistry, Simon Fraser University, 8888 University Drive, Burnaby, British Columbia V5A 1S6, Canada}
\author{U. Rizwan}
\affiliation{Department of Chemistry, Simon Fraser University, 8888 University Drive, Burnaby, British Columbia V5A 1S6, Canada}
\author{J. Smallcombe}
\affiliation{TRIUMF, 4004 Wesbrook Mall, Vancouver, British Columbia, V6T 2A3, Canada}
\author{J.K. Smith}
\affiliation{TRIUMF, 4004 Wesbrook Mall, Vancouver, British Columbia, V6T 2A3, Canada}
\author{K. Starosta}
\affiliation{Department of Chemistry, Simon Fraser University, 8888 University Drive, Burnaby, British Columbia V5A 1S6, Canada}
\author{C.E. Svensson}
\affiliation{Department of Physics, University of Guelph, 50 Stone Road East, Guelph, Ontario  N1G 2W1, Canada}
\author{J. Williams}
\affiliation{Department of Chemistry, Simon Fraser University, 8888 University Drive, Burnaby, British Columbia V5A 1S6, Canada}
\author{M. Williams}
\affiliation{TRIUMF, 4004 Wesbrook Mall, Vancouver, British Columbia, V6T 2A3, Canada}
\affiliation{Department of Physics, University of York, York, YO10 5DD, UK}



\date{\today}

\begin{abstract}
The ``Island of Inversion'' for neutron-rich nuclei in the vicinity of N=20 has become the testing ground {\it par excellence} for our understanding and modelling of shell evolution with isospin. In this context, the structure of the transitional nucleus $^{29}$Mg is critical. The first quantitative measurements of the single particle structure of $^{29}$Mg are reported, using data from the d($^{28}$Mg,p $\gamma$)$^{29}$Mg reaction. Two key states carrying significant $\ell =3$ ($f$-wave) strength were identified at $2.40 \pm 0.10$ ($J^\pi = 5/2^-$) and $4.28 \pm 0.04$ MeV ($7/2^-$). New state-of-the-art shell model calculations have been performed and the predictions are compared in detail with the experimental results. Whilst the two lowest $7/2^-$ levels are well described, the sharing of single-particle strength disagrees with experiment for both the $3/2^-$ and $5/2^-$ levels and there appear to be general problems with configurations involving the $p_{3/2}$ neutron orbital and core-excited components. These conclusions are supported by an analysis of the neutron occupancies in the shell model calculations.
\end{abstract}


\maketitle

\section{\label{intro}Introduction}

Changes in the relative energies of shell model orbits, depending on the neutron/proton balance in the nucleus \cite{Brown}, cause the energy spacings of orbitals to evolve as one goes away from stability and this can therefore change the shell gaps and hence the corresponding magic numbers \cite{Sorlin-Review}. This evolution can be studied most effectively by means of single nucleon transfer reactions.  In particular, the ($d$,$p$) reaction selectively populates states with a significant single particle character and, importantly, allows the spectroscopic strength to be mapped.

The ``island of inversion'' in which the neutron-rich (N$\approx$20) isotopes of Ne, Na and Mg exhibit ground states dominated by cross-shell intruder configurations has garnered much attention since the first measurements of their masses at ISOLDE \cite{Thibault,Campi}. The intruder configurations become energetically favoured owing, in part, to a significant reduction in the energy gap at N=20 between the $1s0d$ and $0f1p$ shells. Importantly, over recent years, this region has become a prime testing ground for our understanding of many of the concepts of shell evolution away from $\beta$-stability, including the development of sophisticated shell-model interactions.

\begin{figure}
\includegraphics[width=\columnwidth]{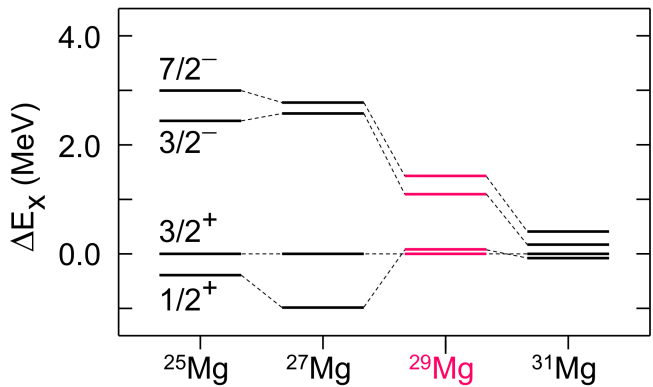}%
\caption{\label{isotope-chains}(Color online)Evolution of intruder state energies for neutron-rich Mg isotopes approaching the island of inversion. The $3/2^+$ level is chosen as the energy reference  (adapted from ref. \cite{Baumann89}). The transitional character of $^{29}$Mg is apparent. }
\end{figure}

One of the keys to understanding the island of inversion lies in the evolution of the energies of the neutron orbitals as we move from near
stable nuclei into this region.  In the case of the Mg isotopes, the single-particle structure of $^{29}$Mg is of key importance to probing the transition into the island of inversion (Fig. \ref{isotope-chains}).  The object of the present work is, therefore, to investigate the $^{28}$Mg($d$,$p$)$^{29}$Mg reaction which permits the transfer of a neutron into the $0d_{3/2}$, $0f_{7/2}$, $1p_{3/2}$ and higher lying orbitals.  As such, the energies of the observed strongly populated (or ``single-particle'') states may be related to the spacing between the neutron $sd$ and $fp$ orbitals.

Very recently, new effective shell model interactions have been developed from first principles (using the Extended Kuo-Krenciglowa (EKK) method \cite{EKK}) and including specifically three-body forces \cite{EEdf1-Tsunoda}.  The effective interaction designated ``EEdf1'', developed for the $sd-pf$ shells \cite{EEdf1-Tsunoda}, has proven capable of reproducing many of the properties of the neutron-rich Ne, Mg and Si isotopes and has
provided new insights into the mechanisms
underlying the related shell evolution and therefore the formation of the island of inversion \cite{EEdf1-Tsunoda}.  These shell model calculations using the EEdf1 interaction have been key to understanding the structure of $^{30}$Mg as studied via intermediate-energy single-neutron removal from $^{31}$Mg \cite{Bea}.
In particular, this work indicated that the transition into the island of inversion is far more gradual and complex than previously thought \footnote{Specifically, the transition into the island of inversion was considered to be very clear between $^{30}$Mg and  $^{31}$Mg.}, and suggested a much more nuanced picture whereby intruder particle-hole configurations ($2p$-$2h$,$4p$-$4h$, $\ldots$)  represent major components of the wavefunctions of the ground and low lying levels.

As indicated above, the most direct means to understand the changes in shell structure in this region -- and indeed to test the new interaction -- is to establish the neutron single particle structure of $^{29}$Mg.

\section{\label{structure}Levels and structure of $^{29}{\rm Mg}$}

The structure of $^{29}$Mg has previously been studied by $\beta \gamma$ coincidences in the $\beta$-decay of $^{29}$Na \cite{Baumann87, Shimoda2014}, by $\beta n\gamma$ coincidences in the $\beta$-decay of $^{30}$Na \cite{Baumann89} by three-neutron transfer using the reactions ($^{11}$B,$^{8}$B) \cite{DavidScott} and ($^{18}$O,$^{15}$O) \cite{Fifield85} with a $^{26}$Mg target, by a multinucleon transfer reaction that adds a single neutron $^{30}$Si($^{13}$C,$^{14}$O)  \cite{Woods88} and by high-energy single-neutron removal from $^{30}$Mg  \cite{RussTerry}. The presently known levels of $^{29}$Mg are summarized in the final columns of Table \ref{exp-sm-levels}. The selectivity observed in the $^{30}$Si($^{13}$C,$^{14}$O)$^{29}$Mg reaction led to the suggestion \cite{Fifield85,Woods88} that the states observed at 1.095 and 1.431 MeV were intruder levels with spin-parity $3/2^-$ and $7/2^-$ respectively. These assignments were consistent with the $\beta$-decay results \cite{Baumann87,Baumann89} and received further support from the $^{30}$Mg neutron-removal experiment where the angular momenta were suggested to be $\ell = 1$ and 3 respectively \cite{RussTerry}, for the removed neutron. The evolution of the energies of the $fp$ intruder states along the Mg isotopic chain is shown in Figure \ref{isotope-chains}. The significance of $^{29}$Mg on the edge of the island of inversion is clear.

The shell model predictions included in the first columns of Table \ref{exp-sm-levels} are from a new calculation using the EEdf1 interaction of ref. \cite{EEdf1-Tsunoda}. This interaction is calculated from a nucleon-nucleon interaction with various computed corrections, and is not fitted to data. The basis for the calculation allowed for cross-shell excitations up to $6\hbar \omega$ for positive parity states and $7\hbar \omega$ for negative parities, which was found to be sufficient for good convergence. The results labelled as {\it wbc} were obtained using the code {\it nushellx} \cite{nushellx, nushellx-RAE, nushellx-MSU} together with a modification of the WBP interaction \cite{WBP} wherein the relative energy of the $pf$-shell was lowered by 0.7 MeV as described in an earlier study  \cite{Brown} of the $^{29}$Mg isotone $^{27}$Ne where this modification was labelled WBP-M. The {\it wbc}, in addition, replaces the USD interaction for the $sd$-shell \cite{USD} with the USD-a interaction \cite{USDA} which is a more appropriate choice in the neutron-rich region. The calculations were restricted to $0\hbar \omega$ for positive parity states as required by the effective interactions, and for negative parity states they included $1\hbar \omega$ excitations from either the $0p$ shell to $1s0d$ or from $1s0d$ to $0f1p$ as described in the original WBP paper \cite{WBP}. The shell model predicts another six states over the next 2 MeV of excitation (with spins of $3/2^-$ and $5/2^-$) that have values of $(2J+1)C^2 S$ between 0.10 and 0.33. These together add to just one unit in $(2J+1)C^2 S$ which means effectively that all observable states up to 6.5 MeV (according to the predictions) are included in the table.

Before reviewing all of the experimental levels, some general comments can be made. A key feature is the pair of $3/2^-$ and $7/2^-$ states just above 1 MeV which represent intruder configurations from the $fp$ shell in which a neutron in the $0f_{7/2}$ or $1p_{3/2}$ orbital is coupled to a $^{28}$Mg core. In this picture, the core can be in its $0^+$ ground state or excited to a higher energy configuration such as $2^+$ but the neutron-transfer reaction can populate these states only via the component with the $0^+$ core, leaving aside any two-step contributions to the reaction mechanism. A second pair of $3/2^-$ and $7/2^-$ states is predicted to lie near 4 MeV in $^{29}$Mg. Of these, the $3/2^-$ is predicted to carry 10-20\% of the single-particle strength that it shares with the 1 MeV partner. The higher lying $7/2^-$ is predicted to carry 30-40\% of the shared single-particle strength. According to the theory, there is evidently a significant mixing between the $7/2^-$ states of $0^+ \otimes 0f_{7/2}$ character and excited-core nature, such as $2^+ \otimes 1p_{3/2}$. There is mixing predicted also between the $3/2^-$ states with $0^+$ and $2^+$ cores. Furthermore, the excited core configurations can include coupling to a neutron in the $0f_{7/2}$ orbital. Another $3/2^-$ state predicted below 3.5 MeV appears not to contain significant single-particle strength relative to the $0^+$ core. Two additional states, each arising from a single excited-core configuration, are the $11/2^-$ state near 3.5 MeV, which arises from $2^+ \otimes 0f_{7/2}$, and the $1/2^-$ state near 2 MeV which arises from $2^+ \otimes 1p_{3/2}$. Of these, just the $1/2^-$ can have a component of single-particle nature with a $0^+$ core, and according to the theory, there is significant mixing and hence an appreciable spectroscopic factor. Finally, the lowest $5/2^-$ state must result from a coupling with an excited core and can mix with the much higher-lying $0^+ \otimes 0f_{5/2}$ configuration, but the mixing is small, at least according to the theory. To summarise, the states built upon excited cores can mix with the $3/2^-$ and $7/2^-$ single particle states and this would result in a significant population of states near 4 MeV that have yet to be identified.

\begin{table*}
\caption{\label{exp-sm-levels}Predicted excitation energies (shell model, present work) and experimental values \cite{NDS29} for states in $^{29}$Mg, together with predicted values of $(2J+1)C^2S$ (which is proportional to the expected transfer cross section) where $S$ is the single-nucleon spectroscopic factor describing $\langle ^{29}{\rm Mg} \mid ^{28}$${\rm Mg} \otimes n \rangle $, $C^2$ is the isospin Clebsch-Gordon coefficient for the (d,p) reaction ($C^2=1$, here) and the transfer is to the $sd-pf$ orbital with the appropriate spin-parity. The neutron separation energy for $^{29}$Mg is $S_n=3.66$ MeV \cite{NDS29}. The list is complete over the range of energies shown, and no further individual states are predicted to have comparable strength up to at least 3 MeV above the separation energy. For shell model details and further discussion, see text.}
\begin{ruledtabular}
\begin{tabular}{c|cc|cc|cc}
$J^\pi $& E$_{\rm{x}}^{\rm{SM}}$ (EEdf1) & $(2J+1)C^2S$ & E$_{\rm{x}}^{\rm{SM}}$ (wbc) & $(2J+1)C^2S$ &E$_{\rm{x}}$ (exp) & Ref. for \\
~ & (MeV) &    &  (MeV)  & ~ & (MeV)  & Assignment  \\ \hline
$3/2^+ _1 $   &        0.000 & 1.41 & 0.090 & 1.61 &0.000   & \cite{Dominique-paper}\\
$1/2^+ _1 $   &        0.026 & 0.70 & 0.000 & 0.79 &0.055   & \cite{Fifield85,Baumann89}\\
$3/2^- _1 $   &        0.872 & 1.66 & 1.350 & 2.50 &1.095   & \cite{Fifield85,Woods88,Baumann87,Baumann89,RussTerry}\\
$7/2^- _1 $   &        1.456 & 3.45 & 1.867 & 3.40 &1.431   & \cite{Fifield85,Woods88,Baumann87,Baumann89,RussTerry}\\
$5/2^+ _1 $   &        1.713 & 0.05 & 1.611 & 0.01 &1.638   & \cite{Baumann87}\\
$1/2^- _1 $   &        1.915 & 0.38 & 2.421 & 0.61 &2.266   & \cite{Baumann89}\\
$5/2^+ _2 $   &        2.106 & 0.26 & 3.147 & 0.33 &3.228   & \cite{Baumann87,Shimoda2014}\\
$3/2^+ _2 $   &        2.129 & 0.77 & 2.269 & 1.00 &2.500   & \cite{Baumann87,Shimoda2014}\\
$7/2^+ _1 $   &        2.195 & $-$  & 2.249 & $-$  &~       & \\
$1/2^+ _1 $   &        2.509 & 0.00 & 2.905 & 0.00 &2.615   & \cite{Baumann87,Shimoda2014}\\
$5/2^- _1 $   &        2.914 & 0.10 & 3.073 & 0.15 &~       & \\
$3/2^+ _3 $   &        2.924 & 0.01 & 3.619 & 0.02 &3.223   & \cite{Baumann87,Shimoda2014}\\
$5/2^+ _3 $   &        3.120 & 0.08 & 3.628 & 0.00 &3.673   & \cite{Baumann87,Shimoda2014}\\
$3/2^- _2 $   &        3.261 & 0.03 & 3.480 & 0.05 &3.090   & \cite{DavidScott,Fifield85,Woods88}\\
$5/2^+ _4 $   &        3.262 & 0.00 & 4.253 & 0.00 &3.985   & \cite{Baumann87,Shimoda2014}\\
$7/2^+ _2 $   &        3.301 & $-$  & 3.992 & $-$  &~       & \\
$11/2^- _1 $  &        3.491 & $-$  & 3.629 & $-$  &~       & \\
$5/2^+    $   &        3.516 & 0.01 & 5.160 & 0.00 &~       & \\
$7/2^+ _3 $   &        3.642 & $-$  & 4.718 & $-$  &~       & \\
$1/2^- _2 $   &        3.767 & 0.66 & 3.646 & 0.83 &~       & \\
$3/2^- _3 $   &        3.832 & 0.42 & 3.973 & 0.34 &~       & \\
$9/2^+ _1 $   &        4.104 & $-$  & 4.077 & $-$  &~       & \\
$7/2^- _2 $   &        4.050 & 1.89 & 4.157 & 2.71 &4.280  & \cite{DavidScott,Fifield85,Woods88} \\
$5/2^- _2 $   &        4.254 & 0.00 & 4.363 & 0.11 &~       & \\
\end{tabular}
\end{ruledtabular}
\end{table*}

The ground state of $^{29}$Mg was deduced to have spin-parity $3/2^+$ \cite{Dominique-paper} on the basis of its decay scheme to known states in $^{29}$Al. This assignment and others for experimentally observed excited states are included in Table \ref{exp-sm-levels}. The
higher energy state in the ground state doublet, at 0.054 MeV, was first proposed to have spin-parity $1/2^+$ by Fifield {\em et al.} \cite{Fifield85} in a re-interpretation of the early $\beta$-decay data \cite{Dominique,Dominique-paper} and this was later confirmed in further $\beta$-decay studies \cite{Baumann89}. The states at 1.095 and 1.431 MeV were postulated \cite{Fifield85} to have spin-parity $3/2^-$ and $7/2^-$ respectively according to the selectivity observed in the $^{30}$Si($^{13}$C,$^{14}$O)$^{29}$Mg reaction. These assignments were consistent with the $\beta$-decay results \cite{Baumann87,Baumann89,Tajiri,Shimoda2014} and the intermediate-energy reaction study mentioned above \cite{RussTerry}. The next higher state at 1.638 MeV was not populated at all in the multinucleon transfer \cite{DavidScott,Fifield85,Woods88} but was observed in the $\beta$-decay of $^{29}$Na (ground state $3/2^+$) and deduced to be $5/2^+$ \cite{Baumann87}. The $\beta$-decay study did not observe the 2.266 MeV state, but did measure and deduce spins and parities for the 2.500 MeV ($3/2^+$) and 2.615 MeV ($1/2^+$) states. These positive parity assignments are supported, where the work overlaps, by a recent study of $\beta$-decay using polarized $^{29}$Na \cite{Shimoda2014}. The 2.266 MeV state was subsequently observed in $\beta$-delayed neutron decay of $^{30}$Na \cite{Baumann89} and was interpreted to have negative parity on the basis of its non-population in the $\beta$-decay of the $3/2^+$ $^{29}$Na ground state; noting also the observed $\gamma$-ray decays (which populate both states in the ground state doublet) and evidence from neutron penetrability arguments, a spin-parity of (1/2,3/2)$^-$ was assigned. The next two states given in the most recent compilation \cite{NDS29} are those at 3.224 MeV and 3.228 MeV that were first observed in $\beta$-decay \cite{Baumann87}. The more recent polarized $\beta$-decay work \cite{Shimoda2014} assigns these as $3/2^+$ and ($5/2$)$^+$ with energies of 3.223 and 3.227 MeV. A level reported in ($^{13}$C,$^{14}$O) at $3.20 \pm 0.04$ MeV \cite{Woods88}, also measured at $3.09 \pm 0.04$ MeV and $3.07 \pm 0.09$ MeV in three-neutron transfer \cite{DavidScott,Fifield85}, was suggested \cite{Woods88} to be a negative parity intruder state. In the compilation \cite{NDS29} this level is associated with the 3.223 MeV $3/2^+$ level, but the interpretation based on the multi-nucleon population \cite{Woods88} suggests that this should be retained as an additional observed state (which is denoted here as 3.090 MeV). Next highest in energy are states at 3.673 MeV and 3.985 MeV that were first observed in $\beta$-decay \cite{Baumann87} and have recently both been assigned as having spin-parity ($5/2^+$) in polarized $\beta$-decay \cite{Shimoda2014}. These two states are above the neutron separation energy of $^{29}$Mg (3.66 MeV). The highest state reported in the compilation is at 4.280 MeV and has previously been seen only in the three multinucleon transfer reactions \cite{DavidScott,Fifield85,Woods88}.

Table \ref{exp-sm-levels} suggests that there are about ten states in $^{29}$Mg predicted by the shell model below 4.3 MeV that are yet to be discovered experimentally. On the other hand, the known experimental states all have reasonable counterparts in the theory.

\section{\label{experimental}Experimental details}

A secondary beam of $^{28}$Mg was obtained from the ISAC2 facility at TRIUMF using a primary beam of 100~$\mu$A of 520 MeV protons bombarding a SiC production target. The extraction of the $^{28}$Mg$^{1+}$ ions was compromised by the failure to hold a sufficiently high voltage on the source and it was necessary to employ a charge state breeder (CSB \cite{Ames2014}) to produce $^{28}$Mg$^{5+}$ ions for injection into the radiofrequency quadrupole at the start of the ISAC acceleration system \cite{Laxdal2014}. The efficiency of the CSB was $10^{-3}$ and it inevitably introduced contaminants. These included radioactive nuclei which had mass to charge ratios close to that of $^{28}$Mg$^{5+}$ and moreover there were stable contaminants derived from the CSB itself. The beam transmitted to the secondary target station comprised $\sim$99\% of the stable isobar $^{28}$Si at a rate of 300,000~pps. Approximately 1\% of the beam, or 3000~pps was found to be the intended isotope $^{28}$Mg ($t_{1/2}=20.9$~h), as discussed below. A smaller amount, estimated as up to $\sim$300~pps, was deduced to be $^{28}$Al ($t_{1/2}=2.2$~m). The energy of the $A=28$ beam was 8.0 MeV/u. The beam spot size on target was $\sim 2$~mm in diameter. The secondary reaction target comprised deuterated polythene (CD$_2$)$_n$ with a thickness of 0.5 mg/cm$^2$.

Elimination of $^{28}$Si-induced reactions from the analysis was achieved using a thin scintillator detector (the \textsc{trifoil}, described below) mounted downstream of the target and preceded by a passive stopper foil. This setup was employed previously \cite{wilson-rutherford} in a similar experiment \cite{wilson-plb} with a radioactive $^{25}$Na beam. In the present work the intensity of the $^{28}$Mg beam was lower than the earlier $^{25}$Na beam by a factor of 10,000 and the mode of operation was different: the passive stopper was used to filter out the higher-Z contaminants, so that only the $^{28}$Mg and $^{29}$Mg reaction products could reach the \textsc{trifoil} and be recorded. The stopper was a 90~$\mu$m thick Al foil. This thickness was sufficient to stop the $^{28}$Si projectiles (and $^{29}$Si reaction products) and to allow all $^{28,29}$Mg ions to reach the \textsc{trifoil} with sufficient energy to be recorded. The Al foil also, as in the earlier experiment \cite{wilson-plb}, stopped any fusion-evaporation reaction products (arising from reactions on the carbon in the target) from reaching the \textsc{trifoil}. The small component of $^{28}$Al in the beam was not anticipated and it was found (see below) that the $^{29}$Al products were able to reach the \textsc{trifoil} in some cases, but only for a particular range of Q-values and only for events with a proton recorded in the backward-most particle detectors.

\begin{figure}
\includegraphics[width=\linewidth]{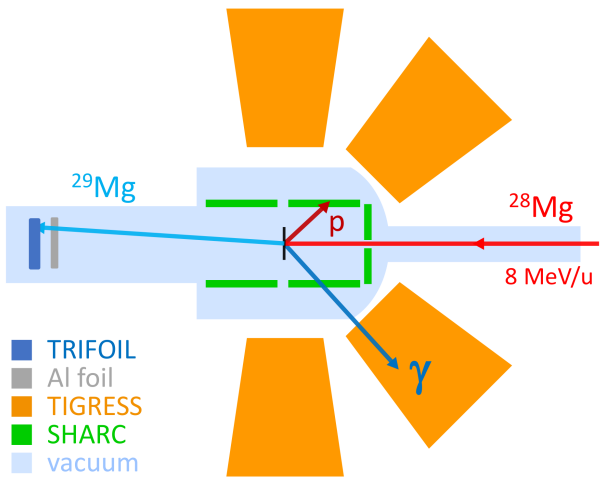}
		\caption{\label{experiment}(Color online) Schematic layout of the experiment, with the beam incident on a deuterated polythene target at the centre of the SHARC silicon strip detector array \cite{SHARC} which is surrounded by 12 TIGRESS clover Ge detectors \cite{TIGRESS} arranged at angles of 90$^\circ$ and 135$^\circ$. Downstream of the target, a passive Al stopper foil prevented fusion-evaporation residues and other contaminant particles from reaching a plastic scintillator detector (\textsc{trifoil}).}
\end{figure}

The experimental setup is shown schematically in Figure \ref{experiment}. The CD$_2$ target was surrounded by the SHARC array \cite{SHARC} which comprises double-sided silicon strip detectors (DSSDs). The downstream box (covering laboratory scattering angles of less than $90^\circ$) was used primarily to detect elastically scattered deuterons for cross section normalisation. The upstream box (laboratory angles from $95^\circ$ to $143^\circ$) and the backward-angle ``CD''annular array (angles $147^\circ$ to $172^\circ$) were employed to record protons from (d,p) reactions.

The \textsc{trifoil} detector was located 400~mm downstream from the target and for the present experiment comprised a square 25~$\mu$m thick BC400 plastic scintillator foil of area $40\times 40$~mm$^2$ aligned axially with the beam. The scintillator was viewed by three photomultipliers and a NIM logic signal was generated if any two photomultipliers responded in coincidence. The reaction angle spanned by the largest circle inscribed within the square scintillator foil was $2.8^\circ$, fully encompassing the $^{29}$Mg products from (d,p) reactions ($<2^\circ$ for protons recorded in the upstream detectors) and elastically scattered $^{28}$Mg particles (for centre-of-mass scattering angles up to $40^\circ$).

Gamma-rays were recorded in the TIGRESS array of HPGe clover detectors \cite{TIGRESS,Hackman2014}, mounted at a distance of 110~mm from the target and operated without any active escape suppression. A total of 12 clovers were deployed, of which 8 were centred at $90^\circ$ and 4 at $135^\circ$ with respect to the beam, spanning all polar angles. An add-back algorithm was implemented to recover the energies for gamma-rays scattered between different crystals within individual clovers. For all gamma-ray events, the segment signal corresponding to the largest energy was assumed to indicate the location of the initial gamma-ray interaction. This allowed the appropriate correction to be applied to the measured energy to account for the Doppler shift arising from the velocity ($\sim 0.10c$) of the emitting beam-like particle.

\begin{figure}
\includegraphics[width=\linewidth]{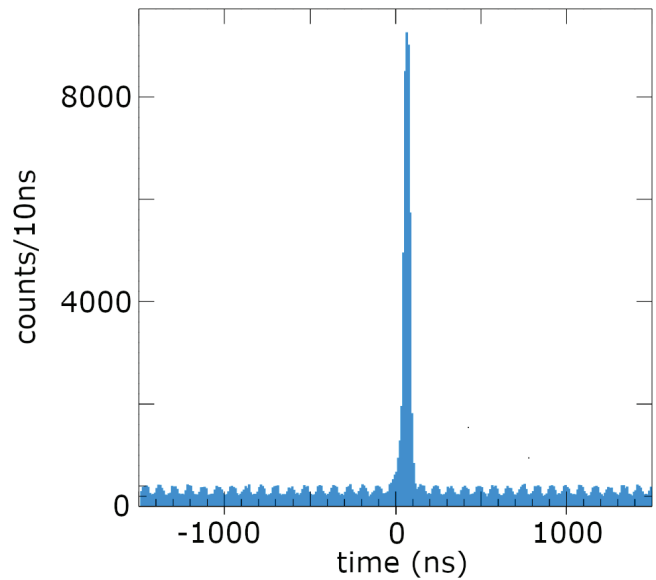}
		\caption{\label{trifoil} (Color online) Spectrum of the relative time between SHARC and \textsc{trifoil} signals for events in which a particle was recorded in SHARC. The main peak corresponds to $^{28}$Mg-induced direct reactions and the small peaks correspond to $^{28}$Mg projectiles by chance being found in nearby beam pulses (see text). By selecting events in a region away from the main peak, a quantitative estimate of the background underlying the peak was obtained. }
\end{figure}

The TIGRESS data acquisition system \cite{TIGRESS-electronics} required a validation (trigger) signal to initiate the data readout. This trigger was derived from the SHARC silicon detectors such that a signal from any strip in SHARC led to the readout of any signals from the \textsc{trifoil} and any coincident silicon and gamma-ray detectors. For the \textsc{trifoil}, the NIM coincidence signal was digitized with a 10~ns sample period over an interval centred on the time of true coincidence pulses. The digital trace was processed to identify signals occurring at the true coincidence time. The adjacent beam pulses, which occurred with a spacing of 86~ns and could easily be distinguished, would also occasionally show signals in the trace if they randomly contained a non-reacting $^{28}$Mg projectile (probability $\sim 3000/(10^9/86)=0.00026=0.026$\%). The times of all logic pulses in the time window were extracted and an example of the relative timing spectrum between the SHARC silicon array and the \textsc{trifoil} is shown in Figure \ref{trifoil}. The subsidiary peaks are also randomly populated when events induced in the SHARC array by the 100-times more-intense $^{28}$Si beam are accompanied by unreacted $^{28}$Mg projectiles in nearby beam pulses (probability $\sim 100 \times 0.00026 = 2.6$\%).

\section{\label{analysis}Analysis}

\subsection{\label{IVa}Overview of analysis}
As described above, it was possible to use the \textsc{trifoil} detector to select the events arising from direct reactions induced by the small $^{28}$Mg component in the beam. In particular these reactions included (d,p), (d,d) and (p,p).  Without the \textsc{trifoil} selection, the kinematic loci for the (d,p) reaction induced by $^{28}$Si projectiles could clearly be observed, along with an underlying background from evaporation protons and alpha-particles. With the \textsc{trifoil} condition imposed, it was clear that the events induced by $^{28}$Si were successfully removed and the kinematic loci corresponding to reactions induced by $^{28}$Mg were observed (Fig.\ \ref{Mg-kinematic}).

The energies recorded in SHARC were assumed to correspond to protons for laboratory angles greater than 90$^\circ$ and deuterons for angles forward of 90$^\circ$ and corrections were applied for the energy losses occurring in the target (assuming reactions at the midpoint) and the dead layers of the silicon detectors. As usual for double-sided silicon strip detectors, the energies recorded on the front and back strips were required to be equal. The position of the beam spot on the target was determined using the observed kinematic line for $^{28}$Si+d elastic scattering as recorded in the various downstream barrel detectors. The d($^{28}$Si,p)$^{29}$Si kinematic lines allowed the positions of the upstream detectors to be fine-tuned. Combined with the known geometry of SHARC, the laboratory scattering angle of the particles recorded in the silicon array could then be determined.

In order to extract absolute cross sections, the integrated luminosity (product of beam exposure and target thickness) was determined using measurements of the deuteron elastic scattering. The differential cross section in counts/msr was first extracted. Since the deuteron energy varies rapidly with the laboratory angle and is measured with good resolution, the energy is the best way to define the scattering angle. Thin cuts in energy were therefore used to define corresponding bins in centre of mass angle. The number of counts in each bin, with suitable background subtraction, was combined with the corresponding solid angle as determined by a Monte-Carlo calculation using {\it Geant4} implemented via {\it NPTool} \cite{NPTool}. In this manner the differential cross section over a range of angles corresponding to $22^\circ$ to $32^\circ$ in the centre-of-mass frame was obtained.

A comparison of the measured elastic scattering cross section in counts/msr with an optical model calculation expressed in mb/sr allowed the luminosity to be deduced. Three optical potentials suitable for this beam-target combination were employed \cite{Daehnick,PereyPerey,Lohr} and these showed a variation between them of 10\% in absolute magnitude over the angular range of interest. The number of counts in each angle bin was determined to an accuracy of 5\%.
The value adopted \cite{Daehnick} for the integrated luminosity was thus ascribed an uncertainty conservatively estimated as 15\%. The analysis of the elastic scattering was validated using the much more intense $^{28}$Si component of the beam (and in fact it was this procedure that gave the best measure of the beam composition, viz.\ 99\% $^{28}$Si and 1\% $^{28}$Mg).

For (d,p) transfer events the energy and angle of the particle observed in SHARC were used, together with the beam energy and assumed reaction kinematics, to calculate the excitation energy of the final nucleus. This procedure was validated using the data for the $^{28}$Si beam which showed peaks in the excitation energy spectrum at the correct energies in $^{29}$Si, including the $1/2^+$ ground state and the strongly populated $3/2^-$ state at 4.93 MeV \cite{Mermaz,Peterson}. In order to derive differential cross sections expressed as mb/msr, the integrated luminosity was taken from the elastic scattering and the solid angle was taken from the calculation using {\it Geant4} and {\it NPTool} \cite{NPTool}.  The differential cross sections were extracted in terms of laboratory angles rather than centre-of-mass angle because it was then possible to identify most clearly the angles that needed to be eliminated due to detector edges, or gaps in the detector coverage, or due to energy detection thresholds.

\subsection{\label{IVb}Results for $^{28}$Mg projectiles}

\begin{table*}
\caption{\label{2j+1c2s}Values of ($2J+1)C^2S$ from fits of ADWA calculations to differential cross sections for the peaks seen in the $^{29}$Mg excitation energy spectrum. Expected states in each region are identified following Table \ref{exp-sm-levels} and the discussion in Section \ref{structure}. The quoted uncertainties are statistical. There are also systematic errors introduced by the peak integration limits ($\pm 0.1$ MeV, corresponding to typically $\pm 10$\% and the normalisation using elastic scattering ($\pm 15$\%, see text). Uncertainties associated with the reaction theory are estimated to be 20\% \cite{Lee}.}
\begin{ruledtabular}
\begin{tabular}{c|cc|cccc||c}
Peak ID & E$_{\rm{x}}$ (min) & E$_{\rm{x}}$ (max) & \multicolumn{4}{c||}{$(2J+1)~C^{2}S$} & Expected \\ \cline{4-7}
 (MeV) &  (MeV)  &  (MeV)  & $\ell = 0$ & $\ell = 1$ & $\ell = 2$ & $\ell = 3$  & states ($J^\pi$) \\ \hline
     0.0    &   $-0.5$ & 0.5 & $0.68 \pm 0.06$ & & $1.20 \pm 0.12$ & ~ & $3/2^+ , 1/2^+ $ \\   
     1.2    & 0.6 & 1.8 &  & $0.44 \pm 0.04$ & & $3.04 \pm 0.16$ & $3/2^- , 7/2^- $ \\ 
     2.4    & 1.9 & 2.9 &  & & $0.32 \pm 0.12$ & $1.80 \pm 0.18$ & $3/2^+ , 5/2^- $ \\ 
     4.2    & 3.7 & 4.7 & & & & $2.40\pm 0.40$ & $7/2^-$ \\ 
\end{tabular}
\end{ruledtabular}
\end{table*}

\begin{figure}
\includegraphics[width=\linewidth]{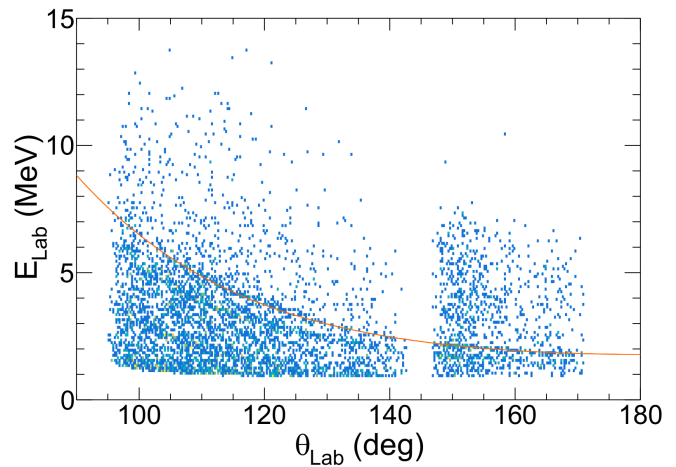}
		\caption{\label{Mg-kinematic}(Color online) Kinematic plot showing proton energy as a function of laboratory angle, after correction for energy losses in the target and in the dead-layer of the silicon detector. The calculated kinematic line for protons populating the ground state of $^{29}$Mg is shown. The origin of the background above this line is discussed in the text.}
\end{figure}

The kinematical plot for the data from $^{28}$Mg projectiles is shown in Figure \ref{Mg-kinematic} for angles backward of $90^\circ$. In order to eliminate low energy signals arising from noise and $\beta$-radiation not eliminated by use of the \textsc{trifoil}, a lower limit was imposed on the detected proton energy (before correction).  The kinematic plot shows a small background of counts above the line corresponding to the ground state of $^{29}$Mg and this is noticeably more intense at angles larger than 145$^\circ$. Whereas the low level of background forward of 145$^\circ$ is explained by the small fraction of $^{28}$Si-induced reactions that escape rejection by the \textsc{trifoil} requirement (owing to random coincidences) the increase at more backward angles has a different origin. This additional background is attributed to a small and unanticipated component (about ten times smaller than the $^{28}$Mg) of $^{28}$Al in the beam. The more positive Q-value for the (d,p) reaction involving $^{28}$Al gives the protons extra energy and they extend to negative excitation energies if the kinematics is assumed to be d($^{28}$Mg,p)$^{29}$Mg.

\begin{figure*}
\includegraphics[width=\linewidth]{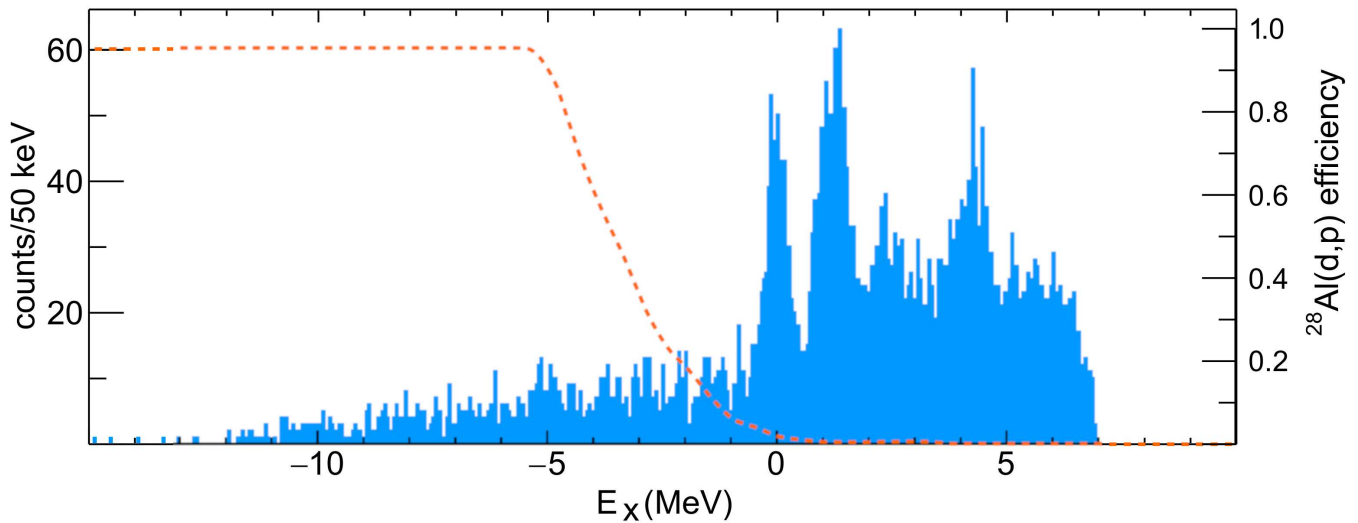}
		\caption{\label{Mg-Ex} (Color online) Excitation energy spectrum for $^{29}$Mg, as deduced from the energy and angle of the proton, for protons having $\theta _{{\rm lab}}>90^\circ$. The background at negative excitation energies is attributed to a small fraction of $^{28}$Al in the incident beam and is calculated to stop at the $^{29}$Mg ground state (see text). The dashed curve shows the probability of any $^{28}$Al-induced (d,p) reaction products being recorded in the \textsc{trifoil} detector according to {\it Geant4} simulations.}
\end{figure*}

Assuming that the events in Figure \ref{Mg-kinematic} correspond to the d($^{28}$Mg,p)$^{29}$Mg reaction, the excitation energy in $^{29}$Mg was computed and is shown in Figure \ref{Mg-Ex}. Fortunately, the {\it Geant4} simulation of the d($^{28}$Al,p)$^{29}$Al reaction shows that the \textsc{trifoil} requirement eliminates any background from this source in the region of positive excitation energies in $^{29}$Mg (cf. Figure \ref{Mg-Ex}). That is, there is an abrupt change in the background at the ground state and the spectrum of  $^{29}$Mg states should therefore have no significant underlying background. In more detail, the simulation also shows that the $^{29}$Al reaction products are only able to reach the \textsc{trifoil} and be recorded if the proton is detected in the CD detector that covers the most backward angles of the SHARC array. This is because the backward-going proton imparts a small extra kick to the forward-going $^{29}$Al ion and also the smaller deflection angle of these $^{29}$Al ions gives them the shortest paths through the passive stopping foil. The clear drop in the background intensity for angles below $145^\circ$ (Figure \ref{Mg-kinematic}) is in excellent agreement with the simulations.

As may be seen in Figure \ref{Mg-Ex}, there are strong peaks observed in the spectrum for $^{29}$Mg at excitation energies of 0.0, 1.2, 2.4 and 4.2 MeV. The possible origins of these peaks are now discussed, keeping in mind that the expected resolution is $\sim$700 keV FWHM (limited mostly by the differential energy loss of protons escaping the target). The peak near 0.0 MeV is likely to contain contributions from both levels comprising the ground state doublet at 0.000 MeV ($3/2^+$) and 0.054 MeV ($1/2^+$). The peak near 1.2 MeV must correspond to the negative parity intruder doublet of 1.095 ($3/2^-$) and 1.431 MeV ($7/2^-$). The peak near 2.4 MeV is open to some speculation, but it does occur close to the known states at 2.266 MeV ($1/2^-$) and 2.500 ($3/2^+$) which can reasonably be expected to be populated (Table \ref{exp-sm-levels}). Additional information from the differential cross sections as discussed below indicates that a previously unobserved negative parity state also contributes. The peak near 4.2 MeV is close to the level reported at 4.28 MeV in multinucleon transfer reactions \cite{DavidScott,Fifield85,Woods88} which was speculated \cite{Woods88} to have negative parity. The asymmetry on the left hand side may point to the population of a weaker state at a slightly lower energy. Interestingly, there is a marked absence of strength at 3.09 MeV where another prominent peak was observed in the multinucleon transfer.

\begin{figure}
\includegraphics[width=\linewidth]{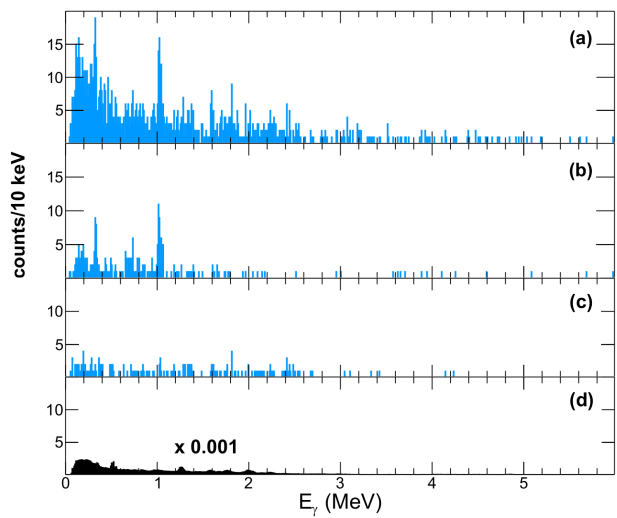}
		\caption{\label{complete-gamma}(Color online) Doppler-corrected gamma-ray energy spectra for $\theta _{{\rm lab}}$(p)$>90^\circ$ : (a) events in which the \textsc{trifoil} is triggered (i.e. mostly corresponding to the $^{28}$Mg-induced (d,p) reaction), (b) with an additional gate of E$_{{\rm x}}$($^{29}$Mg)= $0.8-1.5$ MeV, (c) as for (b) but $2.0-2.8$ MeV, (d) no \textsc{trifoil} gating (i.e. mostly arising from $^{28}$Si-induced reactions). The well-known gamma-rays at 336 keV and 1040 keV from the decay of the 1.431 MeV state are clearly seen in the upper two spectra. Several other tentative peaks from $^{29}$Mg are discussed in the text. }
\end{figure}

The energy spectrum for all gamma-rays recorded in coincidence with SHARC and giving a \textsc{trifoil} signal is shown in Figure \ref{complete-gamma}(a). Clear peaks are observed in the \textsc{trifoil}-gated $^{29}$Mg spectrum, corresponding to the known transitions at 1.095 and 0.336 MeV. It is possible that other peaks occur at several different energies (discussed below) but the limited counting statistics are not conclusive. The spectrum with no \textsc{trifoil} requirement, shown in Figure \ref{complete-gamma}(d), serves to illustrate that any contribution to the \textsc{trifoil}-gated spectrum from the $^{28}$Si projectiles (from both direct and compound reactions) is essentially eliminated.  The gamma-ray energy resolution (after Doppler correction) is 42 keV (FWHM) at 1.095~MeV which is of course far better than the 700 keV (FWHM) resolution for the excitation energy deduced using the protons.

The two states contributing to the peak at 0.0 MeV cannot be distinguished using gamma-rays since the 54 keV transition was not detectable in this experiment (due primarily to the detection threshold and exacerbated by the 1.27 ns lifetime \cite{NDS29} of the state). The gamma-ray energy spectrum for the excitation energy peak near 1.3 MeV is shown in Figure \ref{complete-gamma}(b). The yield of the 1.041 MeV transition exceeds that of the 0.336 MeV by a factor of 3, after correction for efficiency. Given that the 1.431 MeV state decays via a cascade through the 1.095 MeV level, resulting in these two gamma-ray lines, it is clear that both of these states were directly populated in the (d,p) reaction \footnote{Given the limited gamma-ray statistics and the significant lifetime of the 1.431 MeV state ($t_{1/2}=1.4 \pm 0.5$ ns \cite{RussTerry}) a more detailed quantitative analysis was not justified.}.

Unfortunately the gamma-ray statistics for other states are extremely limited and also the experimental spectrum enhances the Compton edge because the add-back is only within each individual clover (this gives an enhancement at $\sim 230$~keV below the full energy peak). There is very tentative evidence in Figure \ref{complete-gamma}(a) for peaks near 1.6, 1.8, 2.4 and 3.2 MeV. The tentative 2.4 MeV is the highest energy seen in the spectrum in Figure \ref{complete-gamma}(c) gated on E$_{\rm x}=2.0-2.6$~MeV, along with weak indications of a 1.0 MeV peak. It may be that the 3.2 MeV peak is associated with decays to either or both of the ground state doublet by a state near 3.2 MeV that could not be clearly discerned in the proton spectrum of Figure \ref{Mg-Ex}. Similarly the 1.6 and 1.8 MeV gamma-rays could arise in part from a gamma-ray decay branch of the unbound states making up the 4.2 MeV peak.

As it was impossible to select individual states by gating on gamma-ray energy, the differential cross sections d$\sigma$/d$\Omega$ of the four prominent peaks in Figure \ref{Mg-Ex} have been extracted. There was no reliable way to fit a smooth underlying background in the excitation energy spectrum but, on the other hand, the background evident at negative excitation energies should not extend into the positive energy region (as discussed above) and the only other background that should be present would arise from weakly populated states that lie near to the strongly populated states. To the extent that the strongly selected states very much dominate the yield (which is discussed again, at the end of the analysis), it was possible to use the simple integrated number of the counts in each peak over the relevant range of energies. In view of the resolution (FWHM) of 700 keV, a range of 1.0 MeV was generally adopted as shown in Table \ref{2j+1c2s}. As discussed above, the peak near 1.2 MeV is known to comprise the two states at 1.095 MeV and 1.431 MeV, separated by 0.336 MeV and hence this gate was widened to 1.2 MeV so as to include as much as possible of both contributions without extending into other adjacent peaks. The region near 3 MeV appears to contain contributions from several less-strongly populated states, but the limitations of the statistics preclude any quantitative analysis.

The angle bins were chosen to be $4^\circ$ in width (in the laboratory frame) and spanned the angles from $96^\circ$ to $172^\circ$, excluding those from $136^\circ$ to $148^\circ$. This avoided the angles at which the solid angle acceptance was varying rapidly and might be incorrectly calculated if there were small residual misalignments in the setup. Expressed in terms of centre-of-mass angles, the range spanned was approximately $2^\circ$ to $40^\circ$ depending on the excitation energy. The peak near 4.3 MeV required a modified procedure, because it is clear in Figure \ref{Mg-kinematic} that the protons fall below the energy threshold for the largest laboratory angles and hence it was possible only to use the angle bins from $96^\circ$ to $116^\circ$.

\begin{figure}
\includegraphics[width=\linewidth]{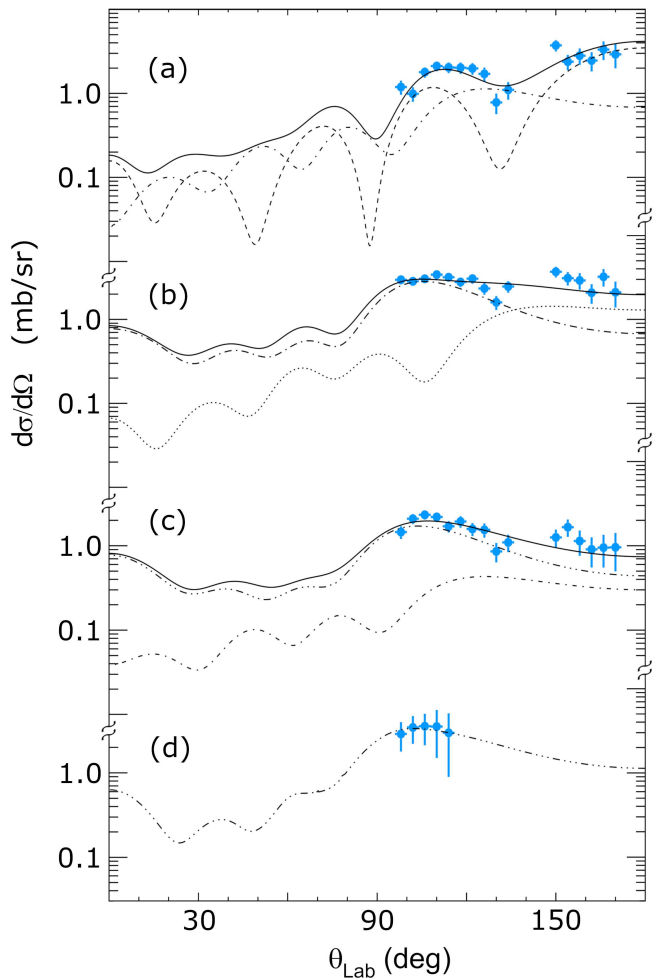}
\caption{\label{angdis-assembly} (Color online) Differential cross sections for the four main peaks identified in the excitation energy spectrum and listed in Table \ref{2j+1c2s}, solid lines are the sum of the different contributions: (a) 0.0 MeV dashed ($\ell=0$, $S=0.34$) and dot-dashed ($\ell=2$, $S=0.30$), (b) 1.2 MeV  dotted ($\ell=1$, $S=0.11$) and dot-dashed ($\ell=3$, $S=0.38$), (c) 2.4 MeV dot-dashed ($\ell=2$, $S=0.08$) and dash-three-dots ($\ell=3$, $S=0.30$), (d) 4.2 MeV dash-three-dots ($\ell=3$, $S=0.30$). }
\end{figure}

In order to determine the angular momentum of the transferred neutron, the differential cross sections were compared with theoretical distributions calculated using the Adiabatic Distorted Wave Approximation (ADWA) of Johnson and Soper \cite{Johnson-Soper}. The code TWOFNR \cite{TWOFNR} was used with standard input parameters \cite{Lee} and the Chapel-Hill (CH89) nucleon--nucleus optical potential \cite{CH89}. As may be seen in Figure \ref{angdis-assembly} the angular momenta were well-determined by the data and multiple $\ell$ contributions were employed where necessary. Spectroscopic factors were deduced by normalising the theoretical curves to the data. The results are collated in Table \ref{2j+1c2s} and discussed below.

\begin{figure}
\includegraphics[width=0.95\linewidth]{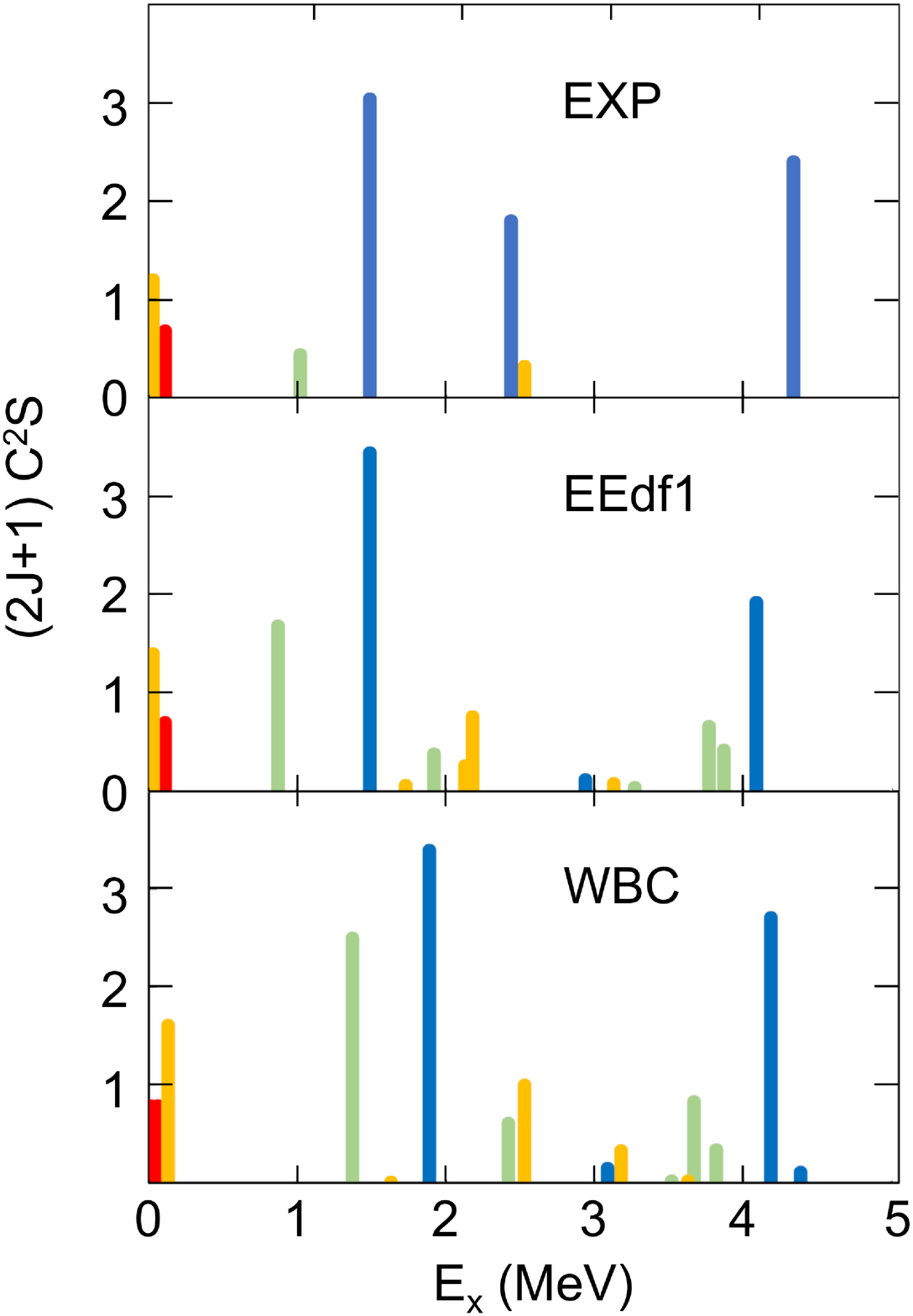}
		\caption{\label{compare-strengths} Comparison of the experimental values of $(2J+1)C^2 S$ and excitation energies from Table \ref{2j+1c2s} with shell model values from Table \ref{exp-sm-levels} (level associations given in Table \ref{SF-table}). Key: (Color online) red $\ell=0$, green $\ell=1$, orange $\ell=2$, blue $\ell=3$.}
\end{figure}

The peak at 0.0 MeV displays  $\ell =2$ and $\ell =0$ contributions to the angular distribution (see Fig. \ref{angdis-assembly}(a)) which is consistent with the known assignments for the $3/2^+$ ground state and the $1/2^+$ first excited 0.054 MeV state, respectively. The distribution (b) for the 1.2 MeV peak is well described by a sum of $\ell = 1$ and $\ell = 3$ contributions, in agreement with the gamma-ray data that indicate the population of both the 1.095 ($3/2^-$) and 1.431 MeV ($7/2^-$) states. The only other known state of similar energy is the 1.638 MeV ($5/2^+$) level \cite{Baumann87,Shimoda2014} and this cannot be populated in single step transfer. The peak at 2.4 MeV in the excitation energy spectrum is less prominent than the other three and hence is the most problematic in the analysis. The differential cross section (c) has a maximum near $105^\circ$ as seen for the peak at 1.2 MeV, and this requires a contribution from $\ell =3$. The behaviour near $180^\circ$ ($0^\circ$ in the centre of mass frame) is slightly different to that in (b), and the fit in this case also demands a contribution from $\ell =2$. The $\ell =2$ component must arise from the level at 2.500 MeV $3/2^+$ if it is from a state that is already known. Regarding the $\ell =3$ component, the only known negative parity state in the region is the 2.266 MeV level that was assigned negative parity in a $\beta$-delayed neutron study \cite{Baumann89}. In the subsequent study of intermediate-energy neutron removal \cite{RussTerry} it was then possible to deduce a spin-parity $(1/2,3/2)^-$. Therefore, the $\ell =3$ strength identified here must correspond to a newly-observed level. From the shell model calculations in Table \ref{exp-sm-levels} the best candidate on the basis of excitation energy is the lowest $5/2^-$ level, predicted according to the EEdf1 calculation at 2.914 MeV. As is clear from Figure \ref{angdis-assembly}(c) the yield in this peak is dominated by the $\ell =3$ state. Hence the peak energy in Figure \ref{Mg-Ex} can be interpreted as the excitation energy of the state, which gives $2.40 \pm 0.10$ MeV. The peak at 4.3 MeV is, unfortunately, observable only for a small range of angles as discussed above. Nevertheless, the distributions shown for $\ell =1$ and $\ell =2$ in (b) and (c) show that the corresponding shapes would give poor descriptions of the data if a single $\ell$-value were dominant. The $\ell =3$ and, less plausibly, $\ell =0$ distributions could account for the data. The two states in the shell model that are consistent with this (cf. Fig. \ref{compare-strengths}) are both populated via $\ell =3$: the $7/2^-$ at 4.050 MeV and the $5/2^-$ at 4.254 MeV. Of these, as shown in the figure, it is only the $7/2^-$ that is predicted to have a strong population in (d,p). While the shell model is under test here, it is reasonable to associate this strong peak near 4.3 MeV with the second $7/2^-$ level. The excitation energy for this level is determined from the spectrum of Figure \ref{Mg-Ex} to be $4.30 \pm 0.10$ MeV and it is natural to associate it also with the 4.28 MeV level reported in multinucleon transfer \cite{DavidScott,Fifield85,Woods88} and listed in the compilation \cite{NDS29}. This level lies above the neutron separation energy, but the experimental resolution is such that it is not possible to set any useful limits on the natural width. In the ADWA calculation for this state, the form factor was derived by assuming a small positive binding energy (and the inferred spectroscopic factor was not sensitive to the precise value).

\begin{figure}
\includegraphics[width=\linewidth]{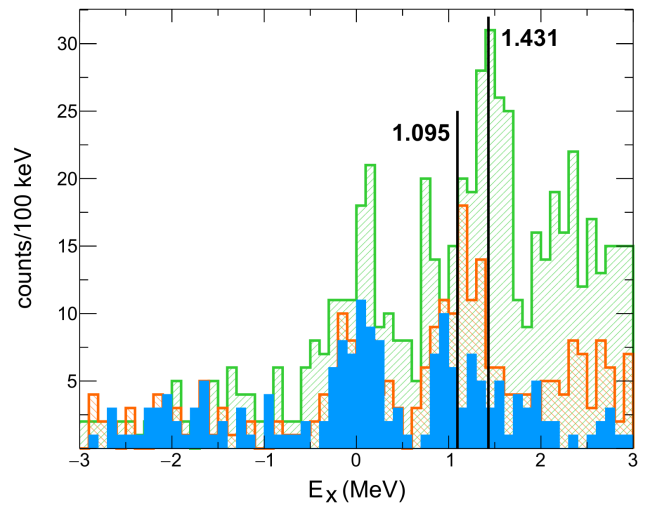}
		\caption{\label{three-angles} (Color online) Excitation energy spectra for $^{29}$Mg corresponding to three restricted angular ranges for protons. {\it Green cross hatched:} $100^\circ$-$110^\circ$, {\it Red cross hatched:} $125^\circ$-$135^\circ$ and {\it Blue solid fill:} $160^\circ$-$170^\circ$. The number of counts is not corrected for solid angle, which varies sinusoidally with angle and is weighted approximately as 7:6:2 for the three spectra. }
\end{figure}

The doublet at 1.2 MeV can be examined in more detail. Although the two contributions are unresolved, they are separated by half of the FWHM for an isolated peak, so the distribution of counts within the energy window can be explored for angle-dependent effects. Three angular ranges were chosen, each of width $10^\circ$ to contain reasonable statistics: $100^\circ$-$110^\circ$, $125^\circ$-$135^\circ$ and $160^\circ$-$170^\circ$. According to the best fit displayed in Figure \ref{angdis-assembly}(b), the state populated with $\ell =3$ should clearly dominate in the first angular range. It should be less dominant in the second angular range, and the $\ell =1$ state should dominate for the third angle (the solid filled spectrum in Figure \ref{three-angles}). It is clear, therefore, that the higher-energy state has $\ell =3$ character and the lower-energy state has $\ell =1$. This then gives the first direct measurement of the orbital angular momenta for these two states and confirms the previous tentative assignments of Refs. \cite{Baumann89,RussTerry}.

\section{\label{discussion}Discussion}

With the spin-parity assignments proposed in section \ref{analysis}, the spectroscopic factors can be deduced from the values of $(2J+1)C^2 S$ presented in Table \ref{2j+1c2s}. These experimental values of $S$ are compared with those predicted by the shell model in Table \ref{SF-table} and the distributions of the strengths $(2J+1)C^2 S$ are compared in Figure \ref{compare-strengths}. There is fairly reasonable agreement, which is discussed in more detail below, but also one notable disagreement. The large value for the spectroscopic factor for the $5/2^-$ level at 2.3 MeV is surprising and is hard to reconcile with the shell model expectations. However, as discussed in detail in section \ref{IVb}, the angular distribution including this level (Fig. \ref{angdis-assembly}(c)) clearly requires a contribution from $\ell =3$. It may be noted that, of the four peaks discussed here, this is the least strongly populated and potentially there could be unidentified background contributions.

\begin{table}
\caption{\label{SF-table}Values of the experimentally deduced spectroscopic factors $S$, using the level identifications discussed in section \ref{analysis}, compared with shell model predictions. The quoted errors in $S$ are statistical. The systematic uncertainties are detailed in Table \ref{2j+1c2s}. Excitation energies are from the literature \cite{NDS29}, cf. Table \ref{exp-sm-levels}, and have experimental uncertainties of $\leq 1$ keV except where indicated (* = present work $\pm 10$ keV; $\ddag$ = $\pm 40$ keV).}
\begin{ruledtabular}
\begin{tabular}{ccc|cc}
E$_{\rm{x}}$ (exp) & $J^\pi _n$ & $S$ & $S$  & $S$   \\
 (MeV) &    & (exp) & (EEdf1) &  (wbc)    \\ \hline
0.000   &   $3/2^+_1$ & $0.30 \pm 0.03 $ &   0.35   & 0.40      \\
0.055   &   $1/2^+_1$ & $0.34 \pm 0.03 $ &   0.35   & 0.40      \\
1.095   &   $3/2^-_1$ & $0.11 \pm 0.01 $ &   0.42   & 0.63      \\
1.431   &   $7/2^-_1$ & $0.38 \pm 0.02 $ &   0.43   & 0.42      \\
2.40*   &   $5/2^-_1$ & $0.30 \pm 0.03 $ &   0.02   & 0.03      \\
2.500   &   $3/2^+_2$ & $0.08 \pm 0.03 $ &   0.19   & 0.25      \\
4.280$\ddag$   &   $7/2^-_2$ & $0.30 \pm 0.05 $ &   0.24   & 0.34      \\
\end{tabular}
\end{ruledtabular}
\end{table}

One of the other striking features of the excitation energy spectrum in Figure \ref{Mg-Ex} is the absence of any strong population of the 3.090 MeV state that dominated the spectra seen in three-neutron transfer \cite{DavidScott,Fifield85}. This state was also populated in the single-neutron transfer (and two-proton pickup) reaction ($^{13}$C,$^{14}$O) \cite{Woods88}. Its most natural association with a shell model state, as shown in Table \ref{exp-sm-levels}, is with the second $3/2^-$ state which has a predicted spectroscopic factor of $S \leq 0.01$. On the other hand, the spectroscopic factors for the overlap of this state with excited core configurations are larger. For the $^{28}$Mg($2_1^+$) core (in the {\it wbc} calculation) these are 0.09 for $2^+ \otimes \nu(0f_{7/2})$ and 0.57 for $2^+ \otimes \nu(1p_{3/2})$. A structure like this would be consistent with the observed strong population of the state in ($^{18}$O,$^{15}$O) and ($^{13}$C,$^{14}$O) -- where the single-neutron transfer could be accompanied by a di-neutron or di-proton transfer with $\ell =2$ -- and also with weak or insignificant population via the (d,p) reaction. The experiment appears to support the predicted lack of mixing between the different $3/2^-$ configurations, but the spectroscopic factor deduced here is significantly smaller than the prediction. In contrast to the situation seen with the first two $3/2^-$ states, there appears to be much more mixing between the $0f_{7/2}$ single particle and  $2^+ \otimes \nu(1p_{3/2})$ configurations so that the first and second $7/2^-$ states each have significant spectroscopic factors for the (d,p) reaction. The spectroscopic factors for the overlap of these states with the $^{28}$Mg($2_1^+$) excited core (in the {\it wbc} calculation) are 0.34  and 0.36 respectively, for $2^+ \otimes \nu(1p_{3/2})$, and rather smaller for $2^+ \otimes \nu(0f_{7/2})$. As such, in both theory and experiment, there is significant single-particle strength in each of these first two $7/2^-$ states.

\begin{table}
\caption{\label{occupancies}Neutron occupancies of the $fp$-shell orbitals according to shell model predictions. For the EEdf1 calculations, the numbers shown are {\it in addition} to the average numbers for the $^{28}$Mg ground state which are shown at the top of the table. The occupancies for {\it wbc} add to slightly less than unity because of excitations from the proton $0p$-shell. The underlined values indicate where the two models differ by more than their overall {\it rms} variation (see text). The spectroscopic factors to the ground state, $S$, are also included.
(* = present work).}
\begin{ruledtabular}
\begin{tabular}{ccc|cccc|c}
E$_{\rm{x}}$ (exp) & $J^\pi _n$ & SM & $\langle n \rangle $ & $\langle n \rangle $  & $\langle n \rangle $ & $\langle n \rangle $ & SM \\
 (MeV) &    & int & $\nu {f_{7/2}}$ &  $\nu {p_{3/2}}$ &  $\nu {p_{1/2}}$ & $\nu {f_{5/2}}$ & $S$ \\ \hline
0.000 & $^{28}$Mg($0^+$) &  EEdf1   & 0.36 & 0.10 & 0.05 & 0.10 & $^{a)}$\\
0.000 & $^{28}$Mg($0^+$) &  wbc     & 0.00 & 0.00 & 0.00 & 0.00 & $^{b)}$ \\ \hline \hline
1.095   &   $3/2^-_1$   &   EEdf1   &   0.17 & \underline {0.55} & 0.01 & $-0.03~$ & 0.42\\
~       &   ~           &   wbc     &   0.23 & \underline {0.72} & 0.03 & 0.01 & 0.63\\ \hline
1.431   &   $7/2^-_1$   &   EEdf1   &   0.53 & 0.22 &  $-0.02~$  & $-0.02~$ & 0.43\\
~       &   ~           &   wbc     &   0.61 & 0.36 & 0.01 & 0.01 & 0.42\\ \hline
2.266   &   $1/2^-_1$   &   EEdf1   &   $-0.04~$ & 0.56 & 0.20 & $-0.02~$ & 0.19\\
~       &   ~           &   wbc     &   0.05 & 0.58 & 0.32 & 0.02 & 0.30 \\ \hline
2.40*   &   $5/2^-_1$   &   EEdf1   &   0.43 & \underline {0.29} & 0.01 & $-0.01~$ & 0.02\\
~       &   ~           &   wbc     &   0.37 & \underline {0.53} & 0.04 & 0.04 & 0.03\\ \hline
~       &   $1/2^-_2$   &   EEdf1   &   0.25 & 0.11 & 0.34 & 0.02 & 0.33\\
~       &   ~           &   wbc     &   0.22 & 0.24 & 0.43 & 0.04 & 0.42\\ \hline
3.090   &   $3/2^-_2$   &   EEdf1   &   0.33 & \underline {0.29} & 0.11 & $-0.02~$ & 0.01\\
~       &   ~           &   wbc     &   0.24 & \underline {0.62} & 0.08 & 0.02 & 0.01\\  \hline
4.280   &   $7/2^-_2$   &   EEdf1   &   0.31 & 0.38 & 0.03 & $-0.02~$ & 0.24\\
~       &   ~           &   wbc     &   0.44 & 0.43 & 0.04 & 0.02 & 0.34\\  \hline
~       &   $5/2^-_2$   &   EEdf1   &   0.26 & \underline{0.46} & \underline {0.03} & $-0.02~$ & 0.00\\
~       &   ~           &   wbc     &   0.25 & \underline{0.16} & \underline {0.48} & 0.08 & 0.02\\
\end{tabular}
\end{ruledtabular}
$^{a)}$The EEdf1 calculation includes excitations up to $4\hbar \omega $ for the $^{28}$Mg g.s.\\
$^{b)}$The wbc calculation requires $0\hbar \omega $ for the $^{28}$Mg g.s.
\end{table}

 The $fp$-shell neutron occupancies predicted in the two shell model calculations are shown in Table \ref{occupancies}. In the case of the EEdf1 results the table gives the {\it excess occupancy} relative to the $^{28}$Mg ground state since the $^{28}$Mg already includes occupation of the $fp$-shell (excitations up to $6\hbar \omega$ or $7\hbar \omega$  are included for positive and negative parity states, respectively). Thus, there is a ``base level'' of excitation into the $fp$-shell (in the EEdf1) that is present in the $^{28}$Mg ground state and which is outside of the WBC basis (and therefore subsumed into the effective interaction). For this reason, we look beyond the inevitable differences between the $^{29}$Mg wave functions as calculated in the two models, and instead focus on comparing the orbitals occupied by the additional neutron in $^{29}$Mg (as given in Table \ref{occupancies}). This highlights those aspects of the wave functions that are most relevant to (d,p) spectroscopic factors. The two calculations are generally in good agreement, with the average difference between the adjusted EEdf1 results and the {\it wbc} being just 0.06 (and the $rms$ difference equal to 0.14). In just five instances the discrepancy exceeds 0.15 and these are underlined in the Table. Intriguingly, all but one of these involve the $\nu$($p_{3/2}$) orbital. Three of the discrepancies concern the two $5/2^-$ wave functions and they reveal differences in the coupling with the excited core, since they occur in the orbitals having a spin different to that of the state. The other two substantial discrepancies concern the two $3/2^-$ states, where the component without any excited core is the source of the disagreement. Interestingly, it is the spectroscopic factors for the $5/2^-$ and $3/2^-$ states that show the largest discrepancy between theory and experiment (cf. Table \ref{SF-table}) as well as between the different theoretical predictions. This indicates that further data for the (d,p) reaction, and in particular a clarification of the $J^\pi$ assignment for the 2.40 MeV state (identified here as the lowest $5/2^-$), would be valuable in distinguishing between the quality of different theoretical predictions and thus refining the models.

 Finally, we note that the aforegoing discussion makes no attempt to address the reduction, or ``quenching’’ of shell model spectroscopic factors that may be expected to arise from effects such as short- and long-range correlations that lie outside the shell model basis \cite{electron, LeeTostevin2006}. The method of analysis employed in the present work has been demonstrated \cite{Tsang2005,Lee} to reproduce (within an accuracy of 20\%) the spectroscopic factors as calculated in conventional large-basis shell model calculations.  Thus, this analysis affords a direct comparison of the experimental results with the theory. A modification to incorporate more realistic bound state wave functions \cite{Kramer2001} -- using, for example, a potential geometry for the bound state wavefunction based on the Hartree-Fock matter density \cite{LeeTostevin2006}
 -- leads to a reduction of around 30\% in the spectroscopic factors deduced from the data. These reduced values show no significant dependence on the nucleon binding energy for isotopes of oxygen \cite{Flavigny2013,Flavigny2018} and argon \cite{Lee2010} and are consistent with the values typically deduced from $(e,e^\prime p)$ scattering \cite{electron}.  Recent results from higher-energy quasi-free knockout, viz.\ two different studies of $(p,2p)$ reactions induced by oxygen isotopes \cite{Atar2018, Kawase2018}, show similar results. Previous studies of nucleon removal reactions at intermediate energies, in contrast, showed a marked dependence of the quenching factor upon the nucleon binding energy \cite{JeffAlex2014} that is not apparent for any other reaction.  None of these effects change in any significant fashion the conclusions of the present work.

\begin{figure*}
\includegraphics[width=0.9\textwidth]{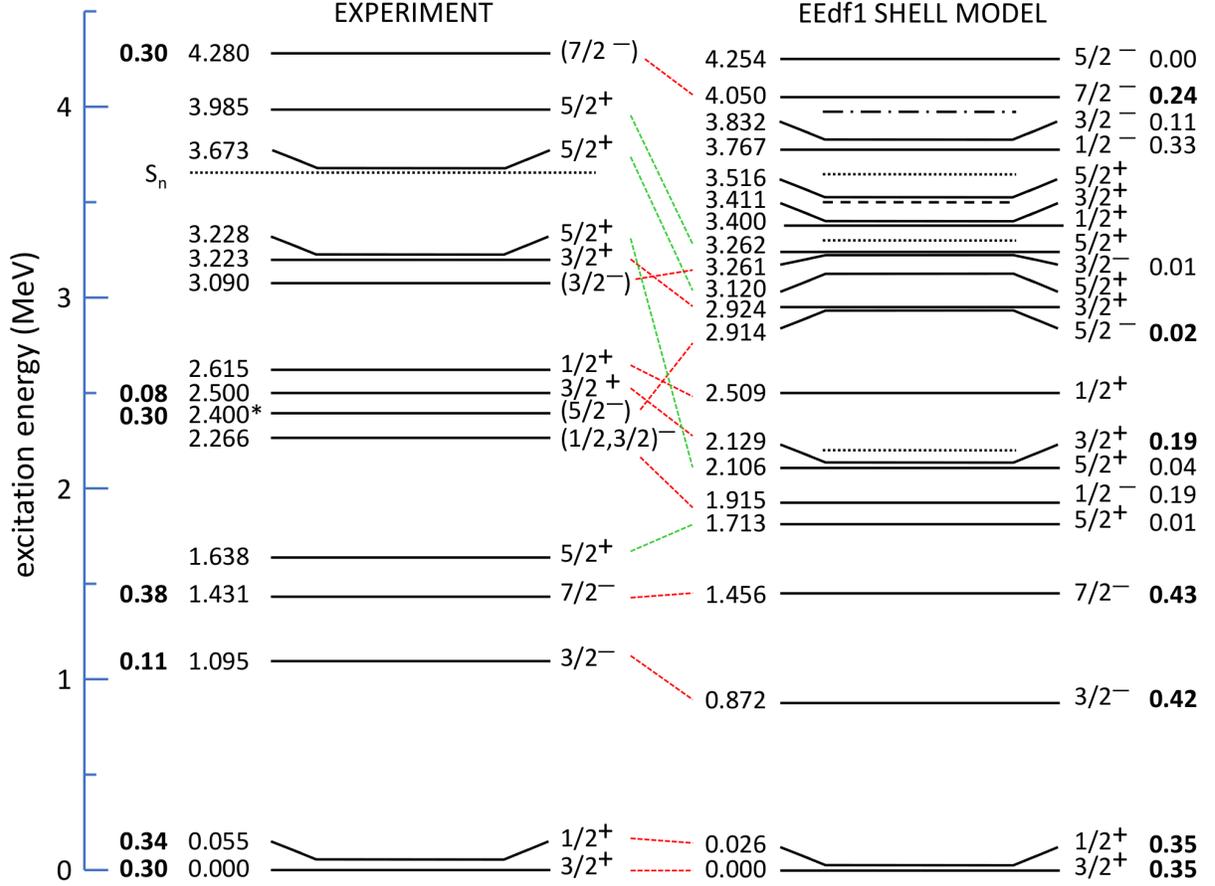}%
\caption{\label{levels}(Color online) Energy levels of $^{29}$Mg. In the experimental level scheme the asterisk (*) denotes the state observed here for the first time and the dotted line shows the neutron separation energy at 3.655 MeV. For clarity, in the shell model level scheme the first $11/2^-$ state is shown by a dashed line, the first $9/2^+$ state by a dot-dashed line and the first three $7/2^+$ levels by dotted lines (note: these five levels, which are included in Table \ref{exp-sm-levels}, cannot be populated in single-step transfer). Other levels are labelled with their spin and parity, excitation energy in MeV and spectroscopic factor $S$. The experimental values of $S$ are from the present work. As shown in Table \ref{exp-sm-levels} the shell model energies for the $5/2^+$ states in the {\it wbc} calculation match better with experiment, whilst keeping the same sequence of spectroscopic factor values as the EEdf1 levels. }
\end{figure*}

\section{\label{summary}Summary and conclusions}

The first quantitative measurements of the single-particle structure of $^{29}$Mg have been obtained using the d($^{28}$Mg,p $\gamma$)$^{29}$Mg reaction. In particular, substantial evidence was found for a previously unknown $5/2^-$ state at $2.40 \pm 0.10$ MeV excitation. Furthermore, considerable $\ell =3$ strength was observed just above the neutron decay threshold in a state at 4.28 MeV that is identified as the second $7/2^-$ level. The present data have also allowed the spins and parities of the two lowest lying intruder states to be confirmed, viz.\  the $3/2^- _1$ at 1.095 and the $7/2^- _1$ at 1.431 MeV. These results offer new insights into the development of nuclear structure approaching the Island of Inversion surrounding $^{32}$Mg.

As summarised in Figure \ref{levels}, and also highlighted in Figure \ref{compare-strengths}, the measurements reveal a marked difference in the spectroscopic strengths associated with the two low-lying negative parity intruder states below 1.5 MeV. This is in contrast to shell model predictions, even though the excitation energies are quite well reproduced. The measurements have also removed the ambiguities that existed in the interpretation of the three-nucleon transfer data \cite{DavidScott,Woods88,Fifield85} and as noted above have located the main part of the remaining intruder strength. As such, the distribution of single-particle strength between the negative parity states appears to be poorly described by the shell model. This is true for both large-basis shell model calculations presented here, despite their very different characteristics. Otherwise, the predicted excitation energies of states and spectroscopic factors for positive parity states are in general in good agreement with experiment.

Whilst the present work has clarified the dominant features of the single-particle structure of $^{29}$Mg, the measurements were compromised by the poor quality and intensity of the $^{28}$Mg radioactive beam. Without the unfortunate factor of 1000 reduction in intensity, the coincident gamma-ray data would have been exploited in the style of Ref.\ \cite{wilson-plb}. In particular, with an effective resolution in excitation energy of some 50 keV, the less strongly populated levels in the region 2.4 to 4.0 MeV could be identified and characterised. Moreover, the gamma-ray decay patterns would provide complementary information concerning the spins of the states.  As such, further measurements using a higher beam intensity would be very worthwhile.


\begin{acknowledgments}
The authors are grateful to the beam operations team at TRIUMF and in particular F. Ames for exceptional efforts in finding some $^{28}$Mg beam, to B. Alex Brown for assistance with the {\it wbc} interaction and the calculations using {\it NuShellX} and to Ahmed Blenz for ensuring that shifts ran smoothly. This work has been supported by the Natural Sciences and Engineering Research Council of Canada (NSERC), The Canada Foundation for Innovation and the British Columbia Knowledge Development Fund. TRIUMF receives federal funding via a contribution agreement through the National Research Council of Canada. The UK authors acknowledge support from STFC. This work was partially supported by STFC Grants ST/L005743/1, ST/P005314/1, ST/L005727/1 and EP/D060575/1. Partial support from the IN2P3-CNRS Projet Internationale de Coop\'eration Scientifique PACIFIC is acknowledged. The shell model work was supported in part by HPCI Strategic Program (hp150224), in part by MEXT and JICFuS and a priority issue (Elucidation of the fundamental laws and evolution of the universe) to be tackled by using Post ``K'' Computer (hp160211, hp170230), in part by the HPCI system research project (hp170182), and by CNS- RIKEN joint project for large-scale nuclear structure calculations.
\end{acknowledgments}

%

\end{document}